\documentclass[sigconf]{acmart}

\usepackage[utf8]{inputenc}
\usepackage{booktabs} 

\copyrightyear{2018}
\acmYear{2018}
\setcopyright{iw3c2w3}
\acmConference[WWW '18 Companion]{The 2018 Web Conference Companion}{April 23--27, 2018}{Lyon, France}
\acmBooktitle{WWW '18 Companion: The 2018 Web Conference Companion, April 23--27, 2018, Lyon, France}
\acmPrice{}
\acmDOI{10.1145/3184558.3192310}
\acmISBN{978-1-4503-5640-4/18/04}

\begin{document}
\title{Learning to Recognize Musical Genre from Audio}
\subtitle{Challenge Overview}

\author{Michaël Defferrard}
\orcid{0000-0002-6028-9024}
\affiliation{%
  \institution{EPFL, Lausanne, Switzerland}
}
\email{michael.defferrard@epfl.ch}

\author{Sharada P. Mohanty}
\affiliation{%
  \institution{EPFL, Lausanne, Switzerland}
}
\email{sharada.mohanty@epfl.ch}

\author{Sean F. Carroll}
\affiliation{%
  \institution{EPFL, Lausanne, Switzerland}
}
\email{sean.carroll@epfl.ch}

\author{Marcel Salathé}
\affiliation{%
  \institution{EPFL, Lausanne, Switzerland}
}
\email{marcel.salathe@epfl.ch}



\begin{abstract}
	We here summarize our experience running a challenge with open data for musical genre recognition. Those notes motivate the task and the challenge design, show some statistics about the submissions, and present the results.
\end{abstract}

\keywords{Music Information Retrieval (MIR); ML Challenge; Open Data}

\maketitle

\section{Introduction}

Like never before, the web has become a place for sharing creative work --- such as music --- among a global community of artists and art lovers. While music and music collections predate the web, the web has enabled much larger scale collections. Whereas people used to own a handful of vinyl records or CDs, they nowadays have instant access to the whole of published musical content via online platforms such as Spotify, iTunes, Youtube, FMA, Jamendo, Bandcamp, etc. Such dramatic increases in the size of music collections has created two challenges: (i) the need to automatically organize a collection (as users and publishers cannot manage them manually anymore), and (ii) the need to automatically recommend new songs to a user knowing their listening habits. An underlying task in both those challenges is to be able to group songs in semantic categories.

Music genres are categories that have arisen through a complex interplay of cultures, artists, and market forces to characterize similarities between compositions and organize music collections. Yet the boundaries between genres still remain fuzzy, making the problem of music genre recognition (MGR) a nontrivial task~\cite{mir_review_genre}. While its utility has been debated, mostly because of its ambiguity and cultural definition, it is widely used and understood by end-users who find it useful to discuss musical categories~\cite{mgr_why}.

The task of this challenge, one of the four Web Conference's challenges, was to recognize the musical genre of a piece of music of which only a recording is available. Genres are broad, e.g.\ \textit{pop} or \textit{rock}, and each song only has one target genre.
Other metadata, e.g.\ the song title or artist name, were not to be used for the prediction.

The data for this challenge comes from the recently published FMA dataset~\cite{fma_dataset}, a dump of the Free Music Archive (FMA).\footnote{\url{https://freemusicarchive.org}}
The dataset is a collection of 917 GiB and 343 days of Creative Commons-licensed audio from 106,574 tracks from 16,341 artists and 14,854 albums, arranged in a hierarchical taxonomy of 161 genres.
It provides full-length and high-quality audio, pre-computed features, together with track- and user-level metadata, tags, and free-form text such as biographies.



\section{The Challenge}

To avoid overfitting and cheating, we organized the challenge in two rounds. The final ranking was based on results from the second round.
The training data for both rounds consisted of the FMA medium subset, which is composed of 25,000 clips of 30 seconds, categorized in 16 genres. The categorization is unbalanced with 21 to 7,103 clips per genre. 
As the data is public, we collected new test data for the second round to prevent access to the test set.

In the first round, participants were provided a test set of 30,000 clips of 30 seconds each and had to upload the predicted genre for each of these clips. The platform readily evaluated those predictions and ranked the participants upon each submission. A subset of these clips were sampled from the FMA large dataset, while ensuring that none overlaps with any clip provided in the training set. The other subset was sampled from songs in the FMA full dataset which are not present in the medium subset.

For the second round, the participants had to provide their models as git repositories which contained the prediction code and the trained model along with an executive summary of their approach.
Docker containers were built out of those repositories.\footnote{\url{https://github.com/jupyter/repo2docker}} We then ran them against a new unseen test set which was sampled from new contributions to the Free Music Archive.

Both rounds used the same evaluation metric. The primary score was the mean log loss and the secondary score was the mean F1 score.
The mean log loss is defined by
\begin{equation}
	L = - \frac{1}{N} \sum_{n=1}^N \sum_{c=1}^{C} y_{nc} \ln(p_{nc}),
\end{equation}
where $N=35000$ is the number of examples in the test set,
$C=16$ is the number of genres,
$y_{nc}$ is a binary value indicating if the $n$-th instance belongs to the $c$-th label,
$p_{nc}$ is the probability according to a submission that the $n$-th instance belongs to the $c$-th label,
and $\ln$ is the natural logarithm.
The mean F1 score is given by
\begin{equation}
	F_1 = \frac{2}{C} \sum_{c=1}^{C} \frac{p^c r^c}{p^c + r^c}, \hspace{1em}
	p^c = \frac{tp^c}{tp^c + fp^c}, \hspace{1em}
	r^c = \frac{tp^c}{tp^c + fn^c},
\end{equation}
where $p^c$ an $r^c$ are the precision and recall for class $c$, and $tp^c$, $fp^c$, $fn^c$ refers to the number of true positives, false positives, and false negatives.

The challenge was hosted on crowdAI, a public platform for open challenges. Instructions on how to participate, access to training and test data, graded submissions, and the leaderboard were available on the challenge page.\footnote{\url{https://www.crowdai.org/challenges/www-2018-challenge-learning-to-recognize-musical-genre}}
Moreover, we developed a starter kit with code to handle the data and make a submission.\footnote{\url{https://github.com/crowdAI/crowdai-musical-genre-recognition-starter-kit}} It also featured some examples and a baseline.
Finally, participants were encouraged to review the FMA paper~\cite{fma_dataset} for a detailed description of the data as well as the GitHub repository for Jupyter notebooks showing how to use the data, explore it, and train baseline models.\footnote{Code and data available at \url{https://github.com/mdeff/fma.}. The \texttt{rc1} version was used.}

\section{Results}


\begin{figure}[t]
\centering
\includegraphics[width=\linewidth]{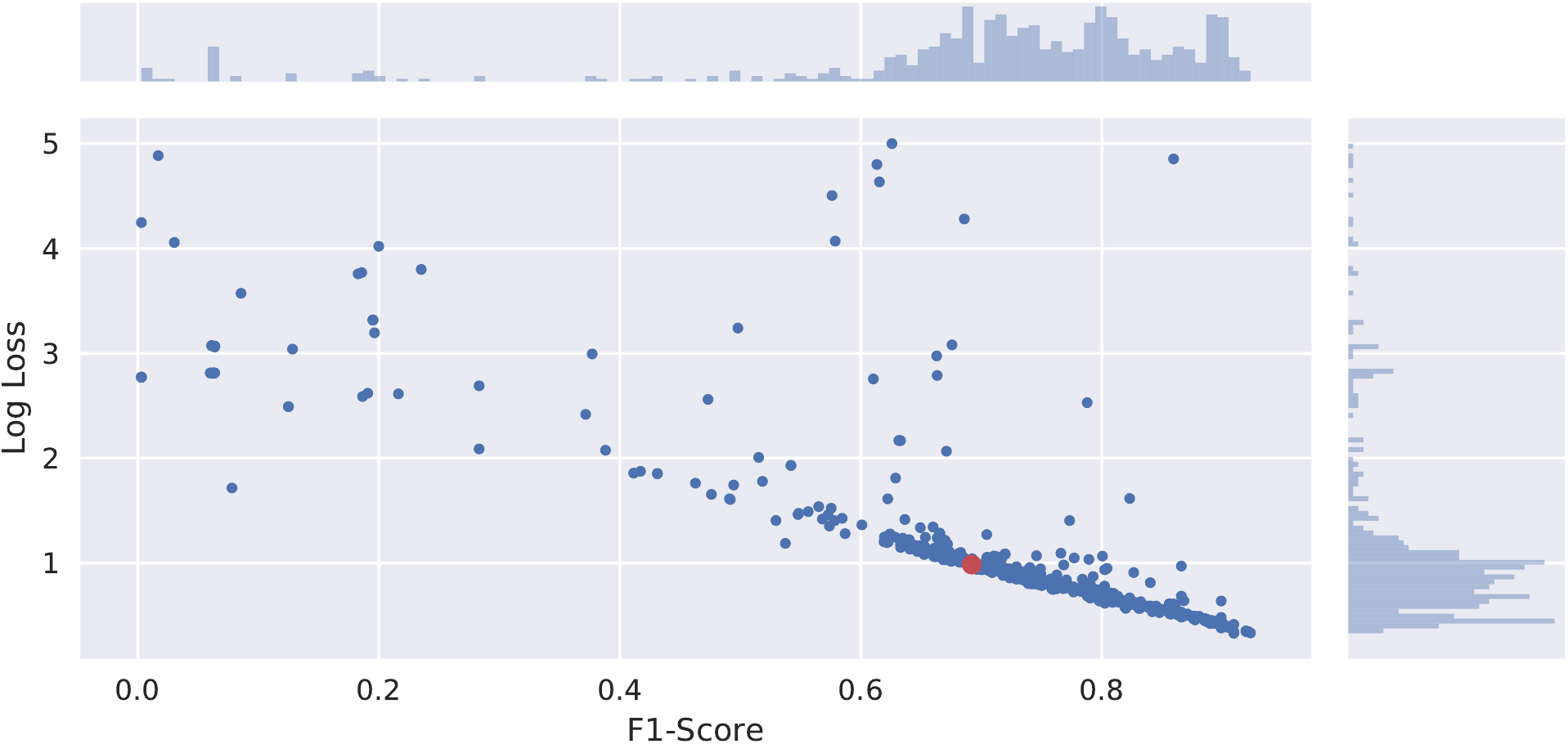}
\caption[dummy]{Joint distribution of the F1 score and the log loss of all submissions.{\footnotemark} The red point represents the baseline prepared by the organizers.}
\label{fig:jointplot}
\end{figure}

\footnotetext{The plot only shows submissions with log loss < 5.}

\begin{figure}[t]
\centering
\includegraphics[width=\linewidth]{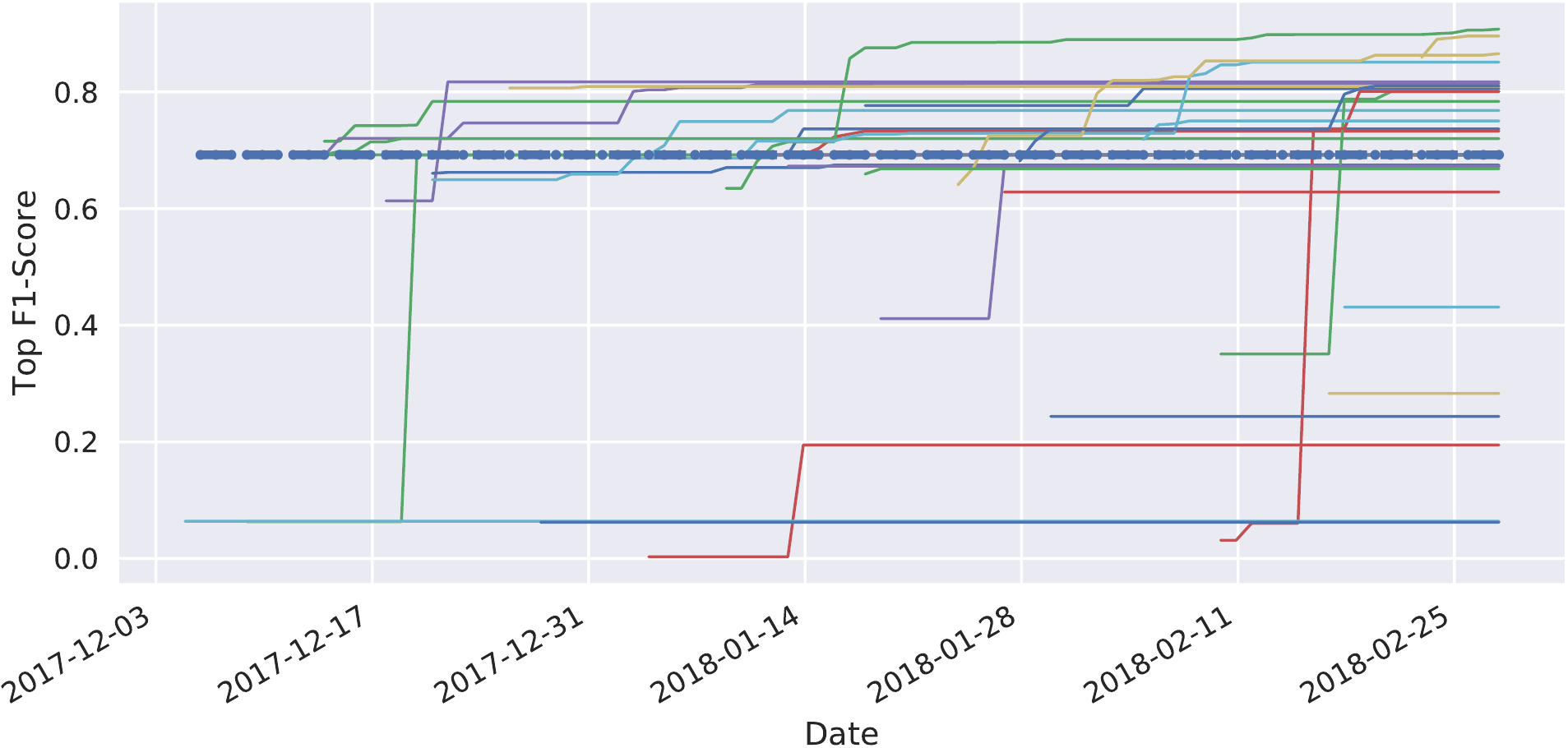}
\caption{Score progression of all participants through the first round of the challenge. Each line represents an active participant.
The dotted line represents the baseline.}
\label{fig:timeline}
\end{figure}

At the end of the first round, we had engaged a total of 246 participants who either made a submission, downloaded the datasets, or contributed to the forums. From those 246 participants, 38 made at least one submission, with some of the top participants making as many as 110 submissions. 
A total of 671 submissions were made in the first round. Of these, 77 were invalid and 576 were successfully graded.
From those 576 submissions, 364 had a score better than the baseline provided by the organizers. Figure~\ref{fig:jointplot} shows the distribution of F1 scores and log losses.
The current best solution has an F1 score of 0.909 and a log loss of 0.330. 
Figure~\ref{fig:timeline} shows how the participants progressed through the first round.

Moreover, we reviewed and accepted two papers.
%
In~\cite{gradient_boosting}, the authors compared the following approaches: (i) ConvNet on spectrograms, and (ii) deep neural net, (iii) ExtraTrees, and (iv) XGBoost on higher-level features extracted by Essentia. They found that ensemble methods outperformed neural networks, with XGBoost performing best.
%
In~\cite{transfer_learning}, the authors argued that genres are subjective and noisy labels, whereas artists are more objective labels.
As an artist is commonly part a subset of genres, and that sets of artists can be seen as exemplars for genres, they hypothesized that musical characteristics which identify an artist may also be key features of certain genres.
As such, they proposed to train a multi-task neural network to jointly predict artist group and genre. Results showed that features learned for artist recognition were indeed useful for MGR.
They thus achieved transfer learning, though keeping the main MGR task as one of the multiple subtasks was crucial.


\section{Conclusion}

The outcomes of the challenge were multiple. First, the accepted papers presented new perspectives and introduced new methods. Then, all the participants to the second round had to share their code as open-source. We hope that those implementations will be useful to the community, for example to serve as baselines, to be scrutinized, or to be improved upon. Finally, the challenge introduced participants to the new FMA dataset and was an opportunity for them to get familiar with it.

That challenge was part of a wider effort to promote open evaluation in machine learning for music data, of which the release of the open FMA dataset was the first step~\cite{fma_dataset}. The goal of this initiative is to establish a reference benchmark based on open data. MIR research has historically suffered from the lack of publicly available benchmark datasets, which stem from the commercial interest in music by record labels, and therefore imposed rigid copyright. The FMA's solution was to aim for tracks which license permits redistribution. All data and code produced during the project and challenge are released under the CC BY 4.0 and MIT licenses.


\bibliographystyle{ACM-Reference-Format}
\bibliography{refs}

\end{document}